# Cryogenic W-band Electron Spin Resonance Probehead with an Integral Cryogenic Low Noise Amplifier

Moamen Jbara,[*] Oleg Zgadzai,[*] Wolfgang Harneit,[#] and Aharon Blank [*,1]

[*] Schulich Faculty of Chemistry, Technion – Israel Institute of Technology, Haifa 3200002, Israel
[#]Universität Osnabrück, Fachbereich Mathematik/Informatik/Physik Institute for physics, Barbarastr. 7, 49076 Osnabrück, Germany

[1] Corresponding author contact details: Aharon Blank, Schulich Faculty of Chemistry, Technion – Israel Institute of Technology, Haifa 3200003, Israel, phone: +972-4-829-3679, fax: +972-4-829-5948, e-mail: ab359@technion.ac.il.

# Abstract

The quest to enhance the sensitivity of electron spin resonance (ESR) is an ongoing challenge. One potential strategy involves increasing the frequency, for instance, moving from Q-band (approximately 35 GHz) to W-band (approximately 94 GHz). However, this shift typically results in higher transmission and switching losses, as well as increased noise in signal amplifiers. In this work, we address these shortcomings by employing a W-band probehead integrated with a cryogenic low-noise amplifier (LNA) and a microresonator. This configuration allows us to position the LNA close to the resonator, thereby amplifying the acquired ESR signal with minimal losses. Furthermore, when operated at cryogenic temperatures, the LNA exhibits unparalleled noise levels that are significantly lower than those of conventional room temperature LNAs. We detail the novel probehead design and provide some experimental results at room temperature as well as cryogenic temperatures for representative paramagnetic samples. We find, for example, that spin sensitivity of $\sim 3 \times 10^5$ spins/$\sqrt{\text{Hz}}$ is achieved for a sample of phosphorus doped $^{28}\text{Si}$, even for sub-optimal sample geometry with potential improvement to $<10^3$ spins/$\sqrt{\text{Hz}}$ in more optimal scenarios.

# I.    Introduction

The quest to enhance the sensitivity of electron spin resonance (ESR) is an ongoing challenge [1]. Nowadays, single electron spin sensitivity has been demonstrated with a variety of detection techniques, such as force [2], electrical [3, 4], single electron transistor [5], nano-superconducting quantum interference device detector [6], direct optical detection [7, 8], indirect optical detection of "dark spins" [9-11], photo-electrical detection [12], and recently also with a microwave fluorescence photon detector [13]. However, despite this wide array of methods, mainstream ESR still primarily uses the induction detection technique, which has far fewer restrictions on sample types, magnetic fields, and measurement temperatures. Induction (Faraday) detection typically employs microwave resonators where the samples are placed. This approach is highly general, works on most samples, provides the highest quality spectroscopic data, and is very efficient when used in conjunction with magnetic resonance imaging methodologies. However, as noted, induction detection is not sensitive enough for many modern applications.

A theoretical analysis of the factors affecting spin sensitivity using conventional induction detection [14-18] reveals that it can be improved by: a) using resonators with the smallest mode volume possible; b) using resonators that have a high quality factor; c) lowering temperatures (as long as this does not cause the spin-lattice relaxation time, $T_1$, to become too long for efficient averaging); and d) increasing magnetic fields (as long as this does not cause spectral broadening, i.e., $T_2^*$, to be too short). Consequently, over the years, many efforts have been invested to improve the sensitivity of induction detection by employing miniature resonators [19-21] (and references therein), extremely high quality factor resonators [22-24], very low temperatures [25], higher magnetic fields [26], and using cryogenic amplifiers [27] (and references therein).

These combined efforts have allowed induction detection to evolve to the point where, with specialized samples that are made as an integral part of the resonator itself, single electron spins could potentially be detected within a reasonable averaging time of a few minutes at millikelvin temperatures [28].

In this work, we aim to push the boundaries of induction detection capabilities of a more general nature by developing a new type of ESR probehead operating at W-band (~94 GHz, ~3.4 T). We present the detailed design, construction method, and testing of the probehead, which includes an integral cryogenic LNA. While cryogenic LNAs have been used in previous ESR designs, including those from our group [21, 25, 27, 29-34], they have not yet been employed at frequencies above ~36 GHz. The new probehead also incorporates newly designed surface microresonators for W-band. These resonators have a mode volume of ~0.1-1 nL, similar to the values obtained in our recent Q-band designs [35, 36], but with potentially higher sensitivity due to their use at higher magnetic fields. Such mode volumes are ~3-4 orders of magnitude smaller than those obtained with the smallest W-band ESR resonators used to date (e.g., loop-gap resonators [37], Fabry-Perot resonators [38], or photonic band gap resonators [39]). The use of microresonators in the present work has two main purposes: a) increasing absolute spin sensitivity, and b) enabling the acquisition of pulse-mode ESR signals under the constraints of limited mm-wave power, as we explain below.

In the following, we provide details of the new cryogenic probehead with an integrated LNA, as well as the dedicated W-band microresonators developed for it. We then describe experimental results obtained using the cryogenic probehead and microresonators on a range of test samples. These results are discussed in the context of achievable ESR spin and concentration sensitivities compared to previous experimental results reported in the literature. Finally, we draw conclusions regarding the current capabilities and potential applications of such cryogenic W-band probeheads for general-purpose pulsed ESR.

## II.     The cryogenic W-band probehead

The rationale behind the design of the new W-band cryogenic probehead is illustrated in Fig. 1. As noted earlier, increasing frequency generally enhances sensitivity, but beyond a certain point, the drawbacks may outweigh the advantages. For instance, when comparing a typical mm-wave W-band (~95 GHz) ESR system to a Q-band (~35 GHz) system, signal loss and amplifier noise increase significantly. Typically, the signal from the resonator must travel approximately 1 meter to reach the mm-wave bridge. This results in losses of about 3.5 dB or more for a WR-10 waveguide [40] and around 1 dB for over-moded structures [41], although additional loss and reflections may occur when adapting back to the WR-10 structure. Furthermore, losses associated with the circulator or any similar transmit/receive decoupling scheme can range from ~1 to ~3 dB. Additional losses come from the protection switch for the LNA, typically 3-4 dB, and the LNA itself, which usually has a noise figure of ~3-6 dB at these frequencies. Ultimately, for room temperature operation, the signal from the resonator is attenuated relative to the noise by approximately 8 dB in the most carefully designed systems and by more than 16 dB in less optimized designs. At cryogenic temperatures, the situation worsens since the original noise levels accompanying the ESR signal should be lower from the outset (depending on temperature). For example, at 30 K, thermal noise power is expected to be 10 times lower than at room temperature, making detection with room temperature noise levels in the LNA highly suboptimal.

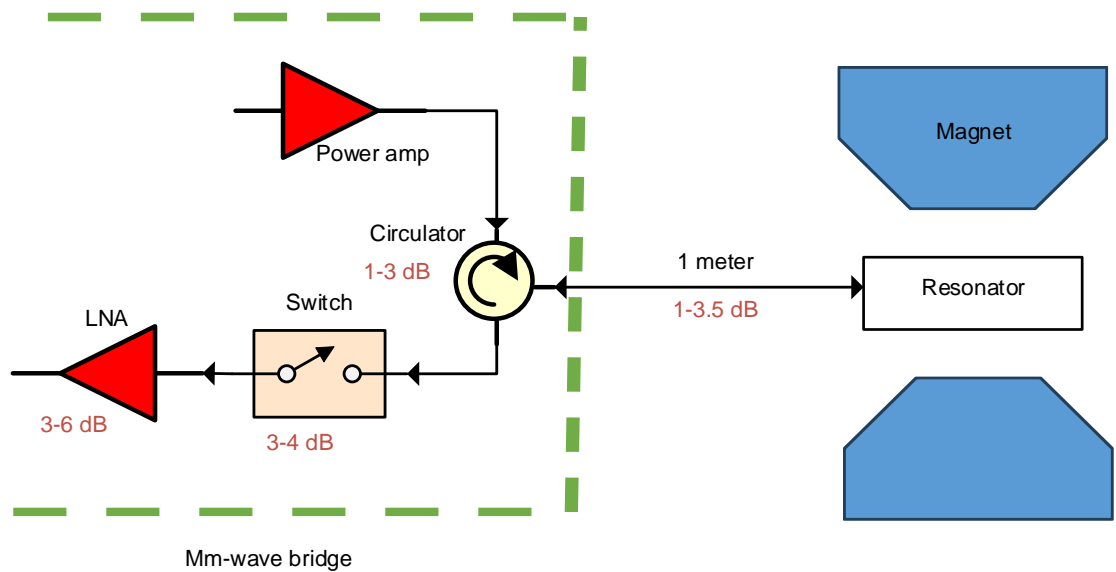

**Figure 1: Typical architecture of the front-end of a mm-wave pulsed ESR system with its corresponding typical loss of signal vs noise level**.

To address these shortcomings, we developed the probehead design shown in Figs. 2-4. The key feature of this design is the placement of a cryogenic LNA (model LNF-LNC65_115WB from Low Noise Factory, Sweden) as close as possible to the resonator, resulting in minimal resonator-to-LNA path loss of approximately 2 dB (measured with a vector network analyzer model N5224B from Keysight, with a frequency extender to W-band). This LNA has a noise figure of ~3 dB at room temperature and a noise temperature of ~25 K when operated at 5 K, representing a significant improvement over conventional LNAs at room temperature with an effective noise temperature of ~440 K (noise figure of 4 dB). Therefore, in terms of minimizing noise in the signal detection chain, our design is nearly optimal. However, this optimized detection chain comes with a trade-off. To fully benefit from the LNA at cryogenic temperatures, it is necessary to attenuate the thermal noise from room temperature components. This is achieved using a directional coupler (model SWD-0640H-10-SB from Eravant, USA, with a 6 dB coupling coefficient) as shown in Fig. 2. The directional coupler replaces the need for a circulator, which typically does not perform well under high static magnetic fields. The Effect of the direction coupler on the noise temperature at the LNA is as follows. For example, if the 6 dB directional coupler is maintained at ~10 K, the noise at its output would be $T_{noise}$ ~ 0.25×300 + 0.75×10 = 82.5 K. For a 10 dB coupler, noise output would be ~39 K. In practice, the noise output is expected to be even lower due to the partial reflection from the resonator and additional losses along the line from the entrance of the probehead, which is cooled to some extent throughout.

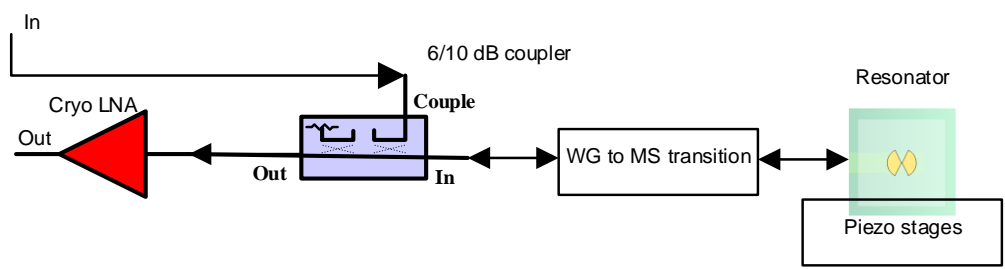

**Figure 2:** **Block diagram of the mm-wave path design of the cryogenic W-band probehead**.

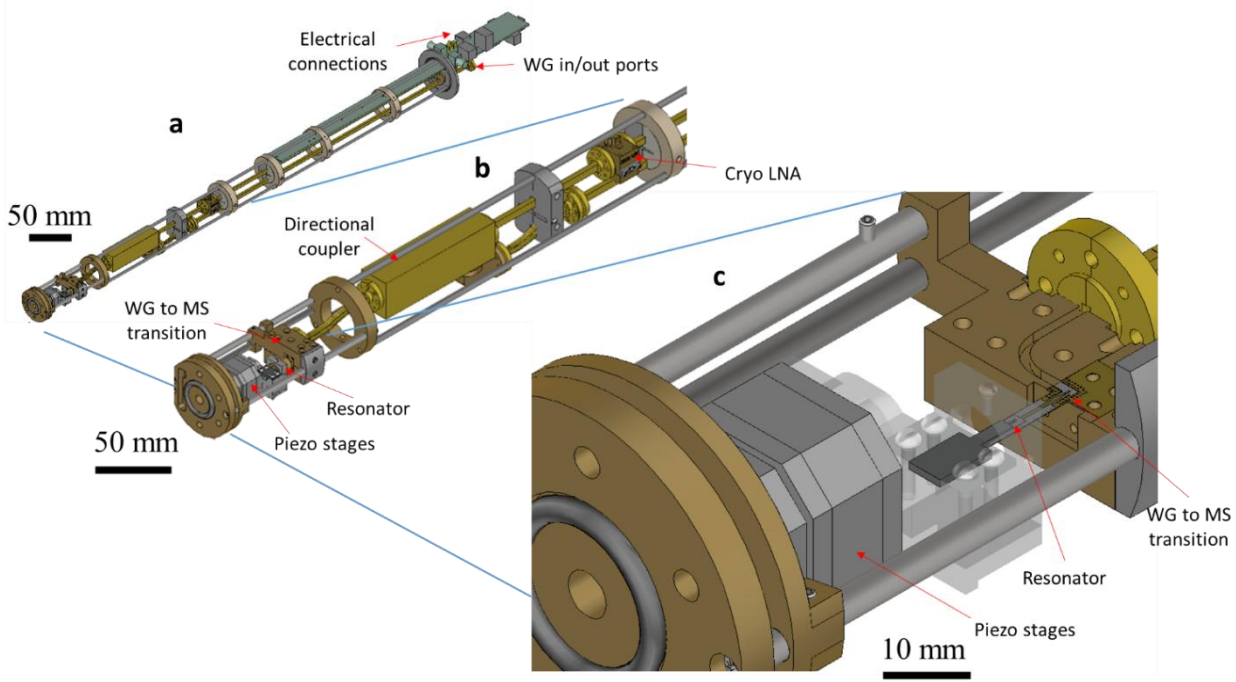

**Figure 3:** **Design of the cryogenic W-band probehead**. (a) Overview of the entire design, featuring two WR-10 waveguide (WG) input and output ports. The probehead is designed to fit a cryostat with a 45 mm inner diameter. (b) Close-up of the distal section of the probehead, highlighting the position of the cryo LNA and the resonator, with coupling controlled by piezo motors. (c) Further zoom-in on the resonator section, showing the WG to microstrip (MS) adapter developed in our previous work [18].

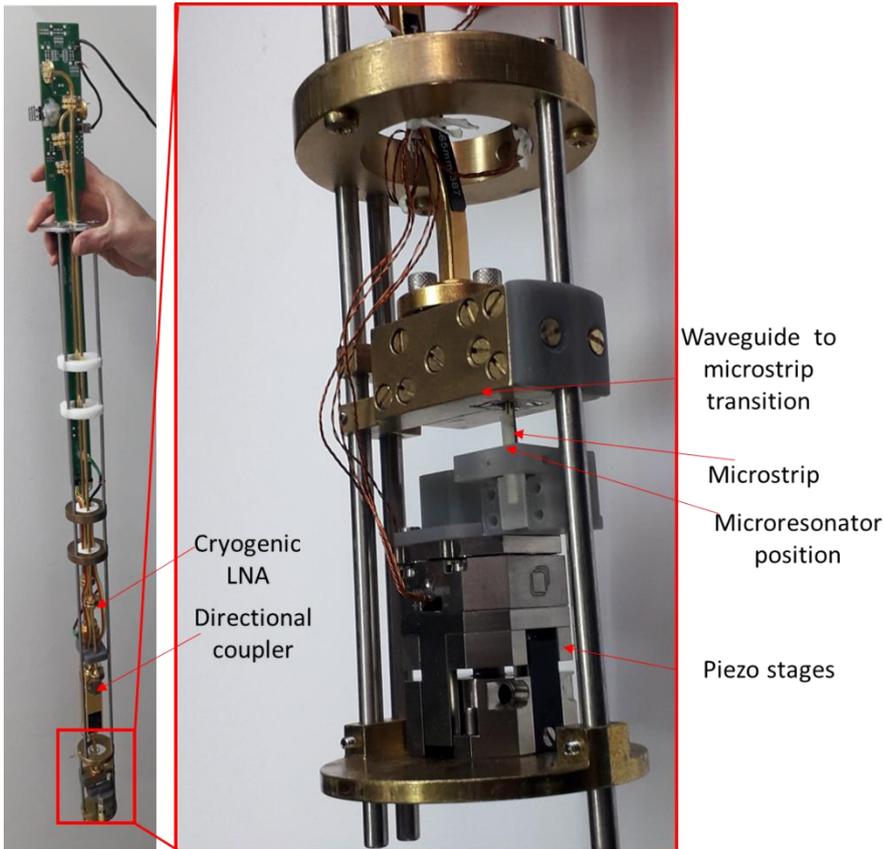

**Figure 4:** **Photos of the assembled cryogenic W-band probehead**. (left) Photo the entire probehead design. (b) Zoom-in to the distal section of the probehead.

Given the above considerations, it is clear that a solution optimizing noise levels in the detection path might prove highly non-optimal for the transmission path, with a required loss level of at least 10 dB. The Appendix discusses these potential issues in more quantitative details. An additional potential shortcoming of our design is the lack of a protection device before the LNA. Such a device (e.g., a PIN diode switch) typically introduces a loss of ~4-5 dB, which would degrade the probehead's performance. Moreover, we were unable to find a waveguide device compact enough to fit into our cryostat. For our selected LNA, input power must be limited to no more than ~0 dBm (1 mW) to avoid damage. This means that, in practice, pulse power reaching the resonator should not exceed 10 dBm, assuming the reflection coefficient ($S_{11}$) of the matched resonator is kept below -10 dB. This limitation restricts the ability to produce short, effective excitation mm-wave pulses. To address this, we turn to the use of surface microresonators, as described in the next section.

## III. W-band surface microresonators

ESR surface microresonators are a relatively recent development in the field and come in several variants [19, 23, 28, 36, 42-46]. In conventional ESR spectroscopy, the resonator's size is typically on the order of the relevant microwave wavelength. For instance, in Q-band ESR, the resonator's typical dimension is around 10 mm. In contrast, ESR microresonators are designed to be much smaller than the wavelength they support, with typical dimensions of $\lambda/100$ to $\lambda/1000$ or less, while still maintaining reasonable quality ($Q$) factors, good spin concentration sensitivity, and excellent absolute spin sensitivity [47-51]. Another key feature of these resonators, especially in the context of this work, is their high mm-wave power to mm-wave magnetic field ($B_1$) conversion ratio, $C_p$. This ratio can reach up to 100 G/√W or more, depending on the exact dimensions, resonance frequency, and $Q$ of the device, enabling efficient pulsed spin excitation with minimal power requirements.

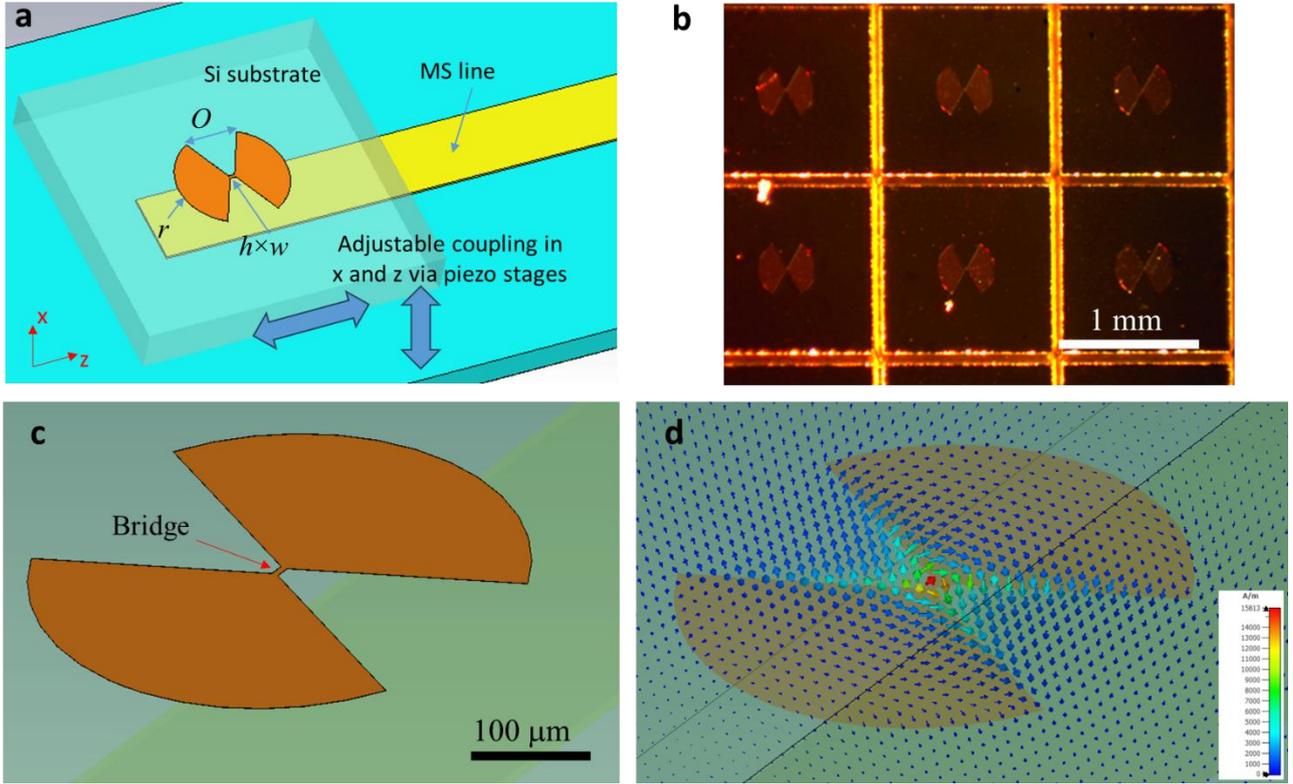

**Figure 5: Surface microresonator of the "ParPar" family for W-band.** (a) General layout of the resonator, consisting of a metallic patch shaped like a butterfly, deposited on a single crystal with high permittivity (silicon in this case). The sample is placed on top of the resonator surface, and resonator coupling is adjusted by changing the relative position of the resonator with respect to the microstrip (MS) line. The dimensions of the resonator are: $r$ – the radius of the metallic patch, $O$ – the opening between two edges as shown, $h$ – the height of the "bridge" section along the MS line, and $w$ the width of the "bridge" section perpendicular to the MS line. (b) Photograph of the resonators fabricated on the silicon wafer. (c) Additional drawing of the resonator and (d) the corresponding calculated mm-wave magnetic field, which is primarily concentrated in the center of the resonator's "bridge" section.

In this work, we employed surface microresonators from the "ParPar" family [36], adapted for operation at W-band (Fig. 5). The resonators were fabricated using photolithography, depositing copper on a thin single-crystal silicon substrate, following the procedure detailed in Appendix II of Ref. [35]. The main characteristics and properties of the resonators used in this study are provided in Table 1. The spin and concentration sensitivities presented in this table were calculated using the expression found in [1] for a sample with $T_1 = 10 \, \mu s$ and $T_2^* = 1 \, \mu s$.

$$Sensitivity_{\sqrt{Hz}}^{spins} \approx \frac{8\sqrt{V_c}\sqrt{k_b T(1/\pi T_2^*)}}{\mu_B \omega_0 \sqrt{2\mu_0}}\sqrt{\frac{\omega_0}{Q_L}}\sqrt{T_1}B_F , \qquad (1)$$

where $V_c$ is the effective volume of the resonator [35], $k_b$ is Boltzmann constant, $T$ is the temperature, $T_2^*$ is the inhomogeneous spins-spin relaxation time, $\mu_B$ is Bohar magneton, $\omega_0$ is the angular resonance frequency, $\mu_0$ is the free space permeability, $Q_L$ is the loaded quality factor of the resonator, $T_1$ is the spin-lattice relaxation time and $B_F$ is the Boltzmann population factor $B_F = \dfrac{1+e^{-\frac{\hbar\omega_0}{k_b T}}}{1-e^{-\frac{\hbar\omega_0}{k_b T}}}$ . Two sample geometries were considered: a small sample that best fits each resonator's dimensions, with dimensions of $h\times h\times w$ mm (where $h$ is the bridge length and $w$ is its width) and placed at the bridge center, and a larger sample that covers almost the entire resonator surface, with dimensions of $200\times200\times50$ µm. For each sample type, we calculated the effective resonator volume [35], from which the absolute spin and concentration sensitivities were deduced. The mode volume, defined as the volume in which 50% of the microwave energy is concentrated [35], is plotted for each resonator in Fig. 6.

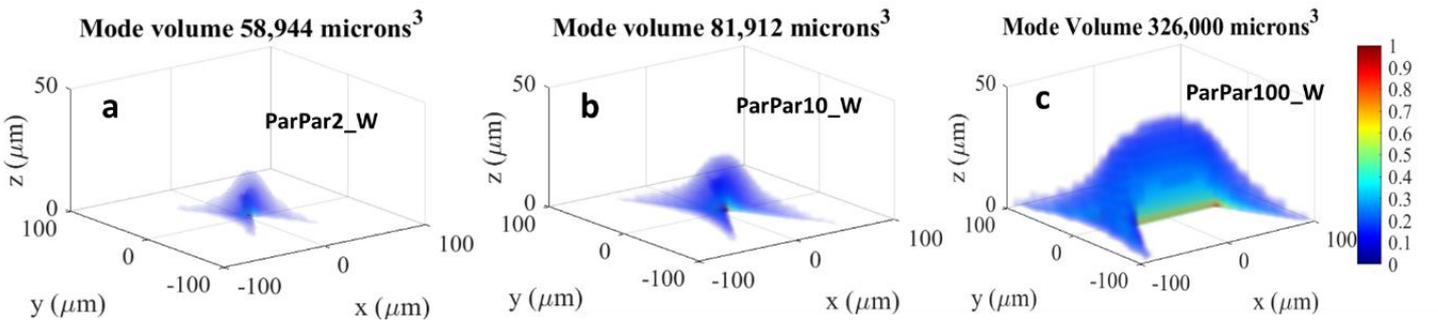

**Figure 6: Calculated mode volume for the surface microresonators used in this work**. The three-dimensional plots represent the volume above the resonator where 50% of the microwave magnetic energy is stored. The color coding indicates the magnitude of the magnetic energy in each voxel, normalized to the voxel with the maximum magnitude of the microwave magnetic field, $B_1$. Plate (a) shows the energy distribution for the ParPar2_W resonator, while plates (b) and (c) display the distributions for ParPar10_W and ParPar100_W, respectively.

| | Structure | | | | Characteristics | | | | Small sample | | | Large sample | | |
|---|---|---|---|---|---|---|---|---|---|---|---|---|---|---|
| **Parameter →**<br>**Resonator type** ↓ | $h$ ($\mu m$) | $w$ ($\mu m$) | $r$ ($\mu m$) | $O$ ($\mu m$) | Res. Frequency, [GHz] | $Q$ ($T$ = 295 100; 10 K) | $C_p$ $\left[\frac{G}{\sqrt{W}}\right]$ | Mode volume [nL] | $V_c$ [ nL] | Absolute spin sensitivity $\left[\frac{spins}{\sqrt{Hz}}\right]$ 295; 100; 10 K. | Concentration spin sensitivity $\left[\frac{nM}{\sqrt{Hz}}\right]$ 295; 100; 10 K. | $V_c$ [ nL] | Absolute spin sensitivity $\left[\frac{spins}{\sqrt{Hz}}\right]$ 300; 100; 10 K. | Concentration spin sensitivity $\left[\frac{nM}{\sqrt{Hz}}\right]$ 300; 100; 10 K. |
| **ParPar2_W** | 2 | 1 | 166 | 170 | 93.8 | 158; 130 ;101 | 628 | 0.059 | 0.0163 | $1.1 \cdot 10^5$; $2.3 \cdot 10^4$; $8.4 \cdot 10^2$ | $4.6 \cdot 10^4$; $9.6 \cdot 10^3$; $3.5 \cdot 10^2$ | 11.64 | $2.88 \cdot 10^6$; $6.12 \cdot 10^5$; $2.24 \cdot 10^4$ | 2.4; $5.1 \cdot 10^{-1}$; $1.87 \cdot 10^{-2}$ |
| **ParPar10_W** | 10 | 5 | 166 | 170 | 92.8 | 95; 93; 92 | 267 | 0.082 | 0.0849 | $3.2 \cdot 10^5$; $6.2 \cdot 10^4$; $2 \cdot 10^3$ | $1.07 \cdot 10^3$; $2.07 \cdot 10^2$; 6.7 | 10.5 | $3.57 \cdot 10^6$; $6.88 \cdot 10^5$; $2.21 \cdot 10^4$ | 2.98; $5.73 \cdot 10^{-1}$; $1.84 \cdot 10^{-2}$ |
| **ParPar100_W** | 100 | 50 | 166 | 170 | 93.1 | 162; 130; 100 | 22 | 0.326 | 8.2 | $2.4 \cdot 10^6$; $5.15 \cdot 10^5$; $1.9 \cdot 10^4$; | 8; 1.72; $6.3 \cdot 10^{-2}$ | 10.1 | $2.65 \cdot 10^6$; $5.71 \cdot 10^5$; $2.09 \cdot 10^4$ | 2.21; $4.76 \cdot 10^{-1}$; $1.74 \cdot 10^{-2}$ |

**Table 1: Main properties of the surface microresonators used in this work.** Resonator dimensions $h$, $w$, $r$, and $O$ are defined in Fig. 5a. The quality factor, $Q$, was measured at various temperatures inside the cryostat using the wide-band tuning mode of the spectrometer, which measures the reflection coefficient as a function of frequency. The value of $C_p$, representing the $B_1$ field for 1 W of mm-wave power, is calculated at a position $w/2$ above the resonator surface (in the rotating frame, with the linear polarization divided by 2). The mode volume is calculated for 50% of the mm-wave magnetic energy (see Fig. 6). The effective volume, $V_c$, is calculated for two cases (with corresponding calculations of spin and concentration sensitivities): first, for a small sample size of $h \times h \times w$ placed at the center of the resonator's surface, and second, for a larger sample size of 200×200×50 μm. The absolute and concentration spin sensitivities are calculated for a hypothetical sample having $T_1 = 10$ μs and $T_2^* = 1$ μs at all temperatures.

# IV  Experimental results

The W-band probehead was tested with a variety of ESR samples at both room temperature and cryogenic temperatures, using the ParPar2_W, ParPar10_W, and ParPar100_W microresonators described above. Measurements were performed with a W-band pulsed ESR spectrometer (SpinUp-W by Spinflex, Israel, with 1 W output power). For each sample type, we measured the spin-lattice relaxation time ($T_1$) using several acquisitions with a variable repetition rate, the spin-spin relaxation time ($T_2$) using the Hahn echo sequence, and estimated $T_2^*$ from the echo time trace. We also quantified the Hahn echo signal-to-noise ratio (SNR) for several transmitted power levels. These measurements were conducted at temperatures of 295 K, 200 K, 100 K, and 10 K, depending on the sample type.

For some samples, we also measured the Carr-Purcell-Meiboom-Gill (CPMG) decay curve and, when possible, used the CPMG sequence to increase the SNR for a given measurement time. All our experiments involved relatively large samples, not confined to the bridge area, so the number of spins actually measured in each sample was estimated based on the resonator mode volume and not the full sample size. In such cases, the expected theoretical SNR was calculated assuming an unoptimized "large" sample size of $200\times200\times50$ µm. Below, we provide more details for each measured sample, along with representative results that demonstrate the capabilities of the new probehead. A summary of the measurement data for different samples, resonators, and temperatures is provided in Table 2.

| Resonator type | Temp. [K] | Sample | Measured sample volume [$nl$] | SNR - Echo | Absolute spin sensitivity $\left[\frac{spins}{\sqrt{Hz}}\right]$ (Calc.) | Spin concentration sensitivity $\left[\frac{nM}{\sqrt{Hz}}\right]$ (Calc.) | SNR – CPMG (# of echoes in the train) | Absolute spin sensitivity $\left[\frac{spins}{\sqrt{Hz}}\right]$ (Calc.) | Spin concentration sensitivity $\left[\frac{nM}{\sqrt{Hz}}\right]$ (Calc.) | $T_1$ [μs] | $T_2^*$ [$ns$] | $T_2$ [$ns$] |
|---|---|---|---|---|---|---|---|---|---|---|---|---|
| ParPar2_W | 10 | $^{28}$Si:P | 0.06 | 353 | $2.11 \cdot 10^6$ ($1.1 \cdot 10^6$)* | 1.76 ($9.1 \cdot 10^{-1}$)* | 1500 (400) | $5 \cdot 10^5$ ($5.5 \cdot 10^4$)* | $4.17 \cdot 10^{-1}$ ($4.3 \cdot 10^{-2}$)* | 970 ± 40 | 150 | 30,000 |
| ParPar10_W | 10 | $^{28}$Si:P | 0.08 | 377.2 | $2.75 \cdot 10^6$ ($1.12 \cdot 10^6$)* | 2.3 ($9.44 \cdot 10^{-1}$)* | 3426 (400) | $3 \cdot 10^5$ ($5.66 \cdot 10^4$)* | $2.5 \cdot 10^{-1}$ ($4.72 \cdot 10^{-2}$)* | 970 ± 40 | 160 | 30,000 |
| ParPar2_W | 200 | N@C$_{60}$ | 0.06 | 130 | $8.75 \cdot 10^7$ ($2.7 \cdot 10^7$)** | $2.43 \cdot 10^3$ ($7.5 \cdot 10^2$)** | - | - | - | 940 ± 40 | 350 | 1725 ± 51 |
| ParPar2_W | 298 | N@C$_{60}$ | 0.06 | 310 | $3.17 \cdot 10^7$ ($1.8 \cdot 10^7$)** | $8.8 \cdot 10^2$ ($5.1 \cdot 10^2$)** | - | - | - | 168 ± 6 | 400 | 2055 ± 173 |
| ParPar10_W | 100 | N@C$_{60}$ | 0.08 | 80 | $1.9 \cdot 10^8$ ($2.2 \cdot 10^7$)** | $3.95 \cdot 10^3$ ($4.66 \cdot 10^2$)** | 21 (10) | $6.25 \cdot 10^8$ ($7.08 \cdot 10^6$)** | $1.3 \cdot 10^4$ ($1.47 \cdot 10^2$)** | 3070 ± 220 | 290 | 813 ± 179 |
| ParPar10_W | 200 | N@C$_{60}$ | 0.08 | 175 | $8.67 \cdot 10^7$ ($4.1 \cdot 10^7$)** | $1.81 \cdot 10^3$ ($8.5 \cdot 10^2$)** | 91 (10) | $1.44 \cdot 10^8$ ($1.28 \cdot 10^7$)** | $3.0 \cdot 10^3$ ($2.66 \cdot 10^2$)** | 940 ± 40 | 215 | 1390 ± 387 |
| ParPar10_W | 298 | N@C$_{60}$ | 0.08 | 400 | $3.28 \cdot 10^7$ ($2.6 \cdot 10^7$)** | $6.83 \cdot 10^2$ ($5.46 \cdot 10^2$)** | 112 (10) | $1.17 \cdot 10^8$ ($8.2 \cdot 10^6$)** | $2.44 \cdot 10^3$ ($1.72 \cdot 10^2$)** | 168 ± 6 | 300 | 2055 ± 173 |
| ParPar100_W | 298 | N@C$_{60}$ | 0.3 | 1100 | $4.47 \cdot 10^7$ ($2.1 \cdot 10^7$)** | $2.48 \cdot 10^2$ ($1.15 \cdot 10^2$)** | 423 (10) | $1.16 \cdot 10^8$ ($6.66 \cdot 10^6$)** | $6.44 \cdot 10^2$ ($36.9$)** | 165 ± 5 | 265 | 2055 ± 173 |

| ParPar2_W | 298 | P1 center | 0.06 | 1220 | $10^9$ $(4.5 \cdot 10^8)$*** | $8.33 \cdot 10^2$ $(3.7 \cdot 10^2)$*** | - | - | - | $1020 \pm 30$ | 125 | - |
|---|---|---|---|---|---|---|---|---|---|---|---|---|
| ParPar10_W | 298 | P1 center | 0.08 | 7000 | $2 \cdot 10^8$ $(5 \cdot 10^8)$*** | $1.67 \cdot 10^2$ $(4.16 \cdot 10^2)$*** | - | - | - | $1020 \pm 30$ | 135 | - |
| ParPar100_W | 298 | Bi-Radical | 0.3 | 235 | $9 \cdot 10^8$ $(3 \cdot 10^8)$# | $5 \cdot 10^3$ $(2.6 \cdot 10^3)$# | - | - | - | $6.2 \pm 0.8$ | 130 | $636 \pm 12$ |
| ParPar100_W | 200 | Bi-Radica | 0.3 | 455 | $4.7 \cdot 10^8$ $(2.7 \cdot 10^8)$# | $2.6 \cdot 10^3$ $(2.3 \cdot 10^3)$# | - | - | - | $14 \pm 1$ | 130 | $758 \pm 10$ |
| ParPar100_W | 150 | Bi-Radical | 0.3 | 390 | $6.1 \cdot 10^8$ $(2.6 \cdot 10^8)$# | $3.4 \cdot 10^3$ $(2.2 \cdot 10^3)$# | - | - | - | $30$## | 130 | - |
| ParPar100 | 100 | Bi-Radical | 0.3 | - | - | - | - | - | - | $48 \pm 10$ | - | - |

**Table 2: Summary of measured data for the new W-band probehead with ParPar resonators.** Data is provided for four types of samples, measured at various temperatures with three types of surface microresonators. For the measured sample volume, if the sample volume placed on the resonator exceeds the 50% mode volume shown in Fig. 6, only the mode volume is considered as the actual measured sample volume. The SNR is given for 1 second of acquisition time, since each sample was measured with a different repetition rate (see text for details). The experimental absolute spin sensitivity is derived from the SNR, measured sample volume, and sample spin concentration. The calculated spin sensitivity (in parentheses) is based on equation (1) [1], assuming 1 second of acquisition with the respective repetition rate for each sample, as noted in the text. For the CPMG train, the SNR may be improved by averaging multiple echoes in the train. The theoretical calculation assumes minimal decay of echoes during the train.

\* The theoretical calculated value takes into consideration that only half of the spins are measured by pulsed ESR (due to the hyperfine interaction).

\*\* The theoretical calculated value takes into consideration that only ~33% of the spins are measured by pulsed ESR (due to the hyperfine interaction).

\*\*\* The theoretical calculated value takes into consideration that only ~20% of the spins are measured by pulsed ESR (due to the hyperfine interaction).

\# The theoretical value takes into account that our pulse excites only ~ 2% of the spins in this sample.

\#\# This value was not measured but was interpolated from the 100 and 200 K results.

a. **Phosphorus doped ²⁸Si:** Phosphorus-doped isotopically enriched ²⁸Si (denoted here as ²⁸Si:P) was used in our experiments. The sample consists of a 50 μm thick ²⁸Si epilayer grown using ²⁸SiH₄ on a Si(100) p-type highly resistive substrate, provided by ISONICS Corporation (USA). The concentration of ²⁹Si in the ²⁸Si epilayer is below 0.1%. The phosphorus concentration is specified at $3.3 \times 10^{16}$ cm⁻³. Echo measurements were performed at a temperature of 10 K with both $\pi/2$ and $\pi$ pulse lengths of 100 ns (the latter being twice the amplitude of the former), a repetition rate of 500 Hz, and using CYCLOPS with ± phase cycling on the first pulse. CPMG measurements were conducted by adding an additional train of 400 $\pi$ pulses, each 100 ns long, with a pulse separation of 700 ns. Figure 7 shows an example of a CPMG train signal obtained for this sample, extending over a duration of more than 250 μs.

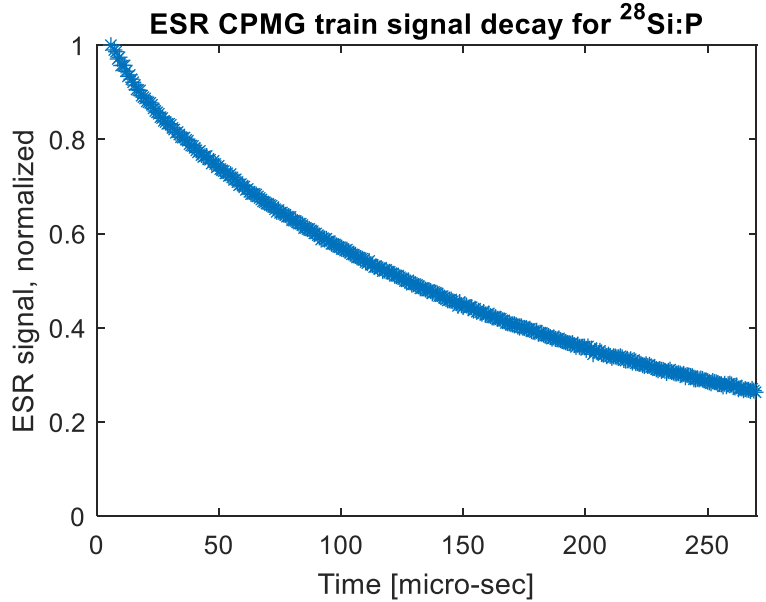

**Figure 7: ESR CPMG train signal.** The signal is acquired every 800 ns for a train of 400 $\pi$ pulses. Effective $T_2$ decay time of this train is ~156 μs.

**b. N@C₆₀:** A powder of N@C₆₀ was prepared using the process described in [52]. The enrichment of the sample, which is the fraction of filled fullerenes, i.e.,

#(N@C$_{60}$)/#(N@C$_{60}$+C$_{60}$), amounts to 310 ppm. The spin concentration, as follows from multiplying this enrichment factor with the molar density of the solid C$_{60}$ crystal, yields ~4×10$^{17}$ spins/cm³. However, for the powder employed in this work, we assume a density that is ~4 times smaller, meaning having in practice ~1×10$^{17}$ spins/cm³. We note that 310 ppm is already a moderately high spin concentration that may lead to dipolar interaction defects like line broadening and/or reduced $T_2$ times. (These effects also occur in 'not-so-dense' solid samples since only the local concentration counts.

The powder was placed on the surface microresonators as shown in Fig. 8a. Three types of microresonators were used with the N@C$_{60}$ samples: ParPar2_W, ParPar10_W, and ParPar100_W. Experiments were conducted at room temperature, 200 K, and 100 K. A typical echo signal and noise (recorded with the static magnetic field off-resonance) measured with the ParPar100_W resonator at room temperature is shown in Fig. 8b. The pulse lengths were 100 ns each, with a pulse separation of 200 ns, and a repetition rate of 10 kHz at room temperature and 600 Hz at both 200 and 100 K, respectively. We also performed CPMG measurements to average multiple echoes before $T_2$ decay, but we were unable to improve the SNR, likely due to rapid $T_2$ decay and pulse sequence imperfections. Additionally, measurements at 200 K and 100 K did not improve the SNR compared to room temperature measurements per averaging time, likely due to an increase in $T_1$ (see Table 2).

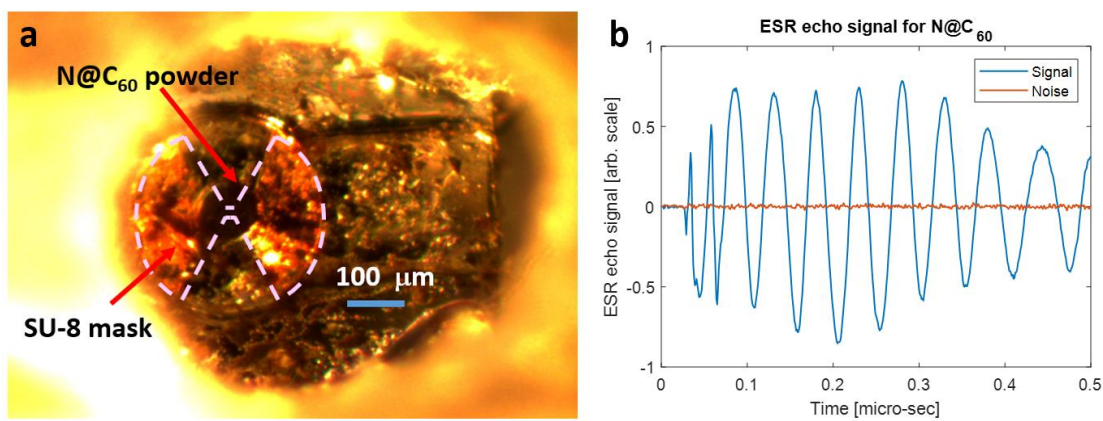

**Figure 8:** **The N@C$_{60}$ sample placed on the surface microresonator and the corresponding ESR signal and noise levels.** (a) Microscopic photo of the powder is placed inside a special round mask, made of SU-8 photoresist by a photolithography process. The powder is held with silicone grease. The outline of the resonator below the sample is marked by a dashed purple line (for ParPar10_W). (b) ESR signal and noise recoded with the ParPar100_W resonator at room temperature. The ESR signal is detected using an off-resonance mixer frequency of 20 MHz away from the carrier for the down conversion stage.

**c. P1 centers in single-crystal diamond:** A high-pressure high temperature (HPHT) single-crystal diamond sample with ~100 ppm of nitrogen content was used in this study (~1.8×10$^{19}$ spins/cm$^3$). The sample was cut to a size of ~200×200×200 μm and was placed on the ParPar2_W and ParPar10_W resonators (see Fig. 9a). The sample exhibits a high ESR signal due to the so-called P1 centers, which are the substitutional nitrogen atoms in the diamond crystal [53]. Experiments were carried out at room temperature. A typical echo signal and noise (recorded with static magnetic field off-resonance), measured with the ParPar10_W resonator at room temperature is shown in Fig. 9b. Pulses length were 100 ns each (with the second pulse twice the amplitude of the first), pulse separation was 200 ns and repetition rate was 1 kHz.

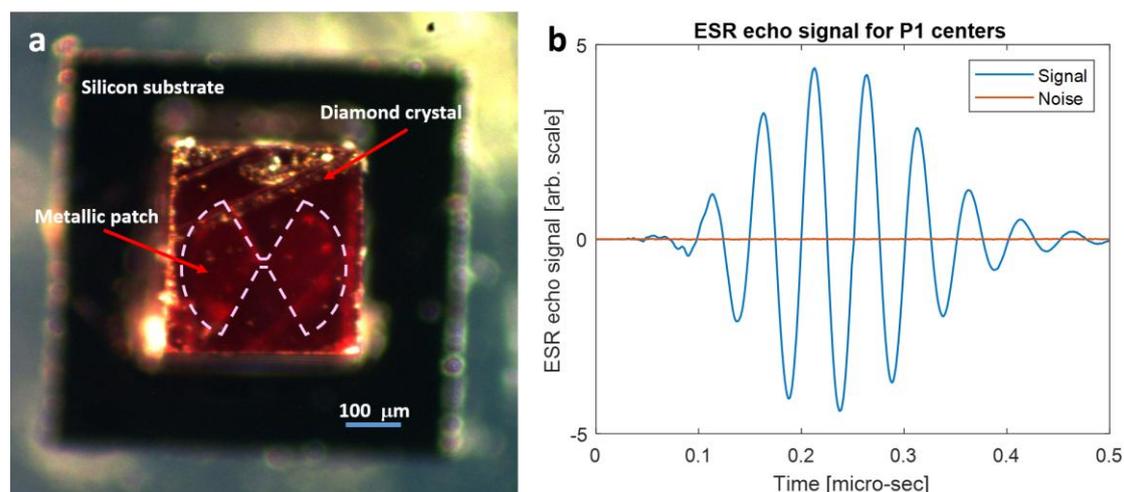

**Figure 9:** **The diamond sample placed on the surface microresonator and its corresponding ESR signal.** (a) Microscopic photo of the diamond sample on ParPar10_W resonator. The outline of the resonator below the sample is marked by a dashed purple line. (b) ESR signal and noise recorded with the ParPar10_W resonator at room temperature. The ESR signal is detected using an off-resonance mixer frequency of 20 MHz away from the carrier for the down conversion stage.

**d. Nitroxide bi-radicals:** Nitroxide bi-radicals are commonly used in double electron-electron resonance (DEER) experiments. In the current experiment, a single grain from a standard solid nitroxide bi-radical sample, provided by Bruker (product #E3005315), was utilized (Fig. 10a). The sample concentration was determined to be approximately $2.4 \times 10^{17}$ spins/cm³ through quantitative CW ESR measurements using a Bruker continuous-wave ESR system. The echo signal was recorded at several magnetic fields and at temperatures of 100 K, 200 K, and 295 K, with the signal-to-noise ratio (SNR) evaluated at each. The field-swept echo signal measured at 295 K is presented in Fig. 10b. The pulse lengths were 150 ns each (with the second pulse twice the amplitude of the first), with a pulse separation of 200 ns and a repetition rate of 100 kHz. Measurements at 200 K were conducted with repetition rate of 50 kHz and at 150 K with 40 kHz (see results in Table 2).

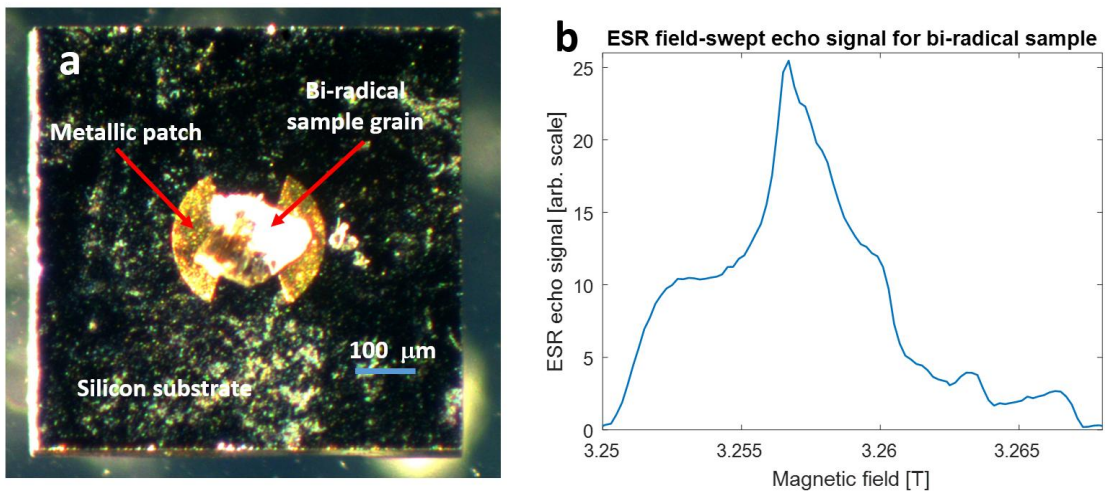

**Figure 10: The solid bi-radical sample placed on the surface microresonator and the corresponding field-swept echo ESR signal.** (a) Microscopic photo of the bi-radical sample on ParPar100_W resonator. (b) The field-swept echo ESR signal recorded with the ParPar100_W resonator at 295 K temperature.

# V Discussion and conclusions

This work presents the first example of implementing a cryogenic LNA within a cryogenic W-band ESR probehead. As a result, we achieved an experimental high spin sensitivity of approximately $3\times10^5$ spins/$\sqrt{\text{Hz}}$ for a $^{28}$Si:P sample at 10 K. For a more conventional sample of solid nitroxide bi-radical, we obtained a spin sensitivity of ~$4.7\times10^8$ spins/$\sqrt{\text{Hz}}$ at 200 K, while for the N@C$_{60}$ sample, the spin sensitivity was ~$3.3\times10^7$ spins/$\sqrt{\text{Hz}}$ at room temperature. It is important to note that these absolute spin sensitivity values were achieved for non-optimized sample geometries. Specifically, the samples were often significantly larger than the resonators' mode volume and were not confined to the areas where the resonators are most sensitive. Another point to consider is that our N@C$_{60}$ sample was highly non-optimized for achieving high spin sensitivity, with $T_2$ much shorter than what could be achieved using other compositions with lower enrichment. While $T_1$ values found for the N@C$_{60}$ sample are approximately 2–3 times shorter than those previously reported for W-band [54], $T_2$ was significantly lower compared to $T_1$. Generally, for sensitivity purposes, it is preferable to reduce $T_1$ to allow for faster averaging, as long as $T_2$ remains unaffected, ideally reaching a state where $T_1 \approx T_2$. This can be achieved, for instance, with a $^{15}$N@C$_{60}$ sample having ~2 ppm of enrichment, as demonstrated previously [55] (with $T_1$~2 ms and $T_2$~200 µs at 70 K). Other emerging spins of interest for improved spin sensitivity include encapsulated atomic hydrogen in octamethyl-POSS cages, which can achieve $T_1$ of ~200 µs and $T_2$ of ~11 µs at 160 K [56].

It should be noted that our present sensitivity figures are already very encouraging and are well beyond the current state-of-the-art for W-band ESR. For example, when using a conventional cylindrical TE$_{011}$ cavity, a sensitivity of ~$3\times10^{11}$ spins/$\sqrt{\text{Hz}}$ was achieved for a sample of 3-Carboxy-TEMPO at 40 K using pulsed ESR (assuming a repetition rate of 10 kHz) [57]. Moreover, experiments carried out with a pulsed millimeter-wave spectrometer at room temperature using a small sample volume (~500 nanoliters) resonator resulted in a sensitivity of ~1.1×10$^{10}$ spins/$\sqrt{Hz}$ for a nitroxide sample [58]. More recent designs achieved a similar level of spin sensitivity of ~2.5×10$^{10}$ spins/$\sqrt{Hz}$ at 115 GHz with a sample of BDPA in polystyrene, at room temperature,[59] and about ~3×10$^{10}$ spins/$\sqrt{Hz}$ at 95 GHz using photonic band gap resonators with ~5 µL of 100 µM aqueous solution of nitroxide Tempol, using CW ESR [39].

While placing a cryogenic LNA directly next to the resonator is highly beneficial for noise reduction and minimizing signal loss, it introduces the challenge of limiting the mm-wave power reaching the resonator. To protect the LNA from damage, we require that the power at the resonator not exceed ~0–10 dBm. For small, geometry-optimized samples located at the center of the resonator's surface, this should not pose a significant problem, particularly for the ParPar2_W and ParPar10_W resonators. In these cases, we expect π pulse lengths of approximately 8 ns and 20 ns, respectively (based on the calculated $C_p$ values in Table 1, assuming 0 dBm at the resonator). However, for the ParPar100_W resonator, this necessitates a π pulse length of around 240 ns, which restricts the ability to conduct measurements that require short pulses, such as in DEER experiments.

Another important aspect to consider is that the choice of resonator size depends on the desired application. Often, absolute spin sensitivity is not the primary focus of the probehead. More frequently, spin concentration sensitivity is the key factor, particularly in applications such as DEER experiments. Concentration sensitivity tends to favor larger resonators [1, 60]; however, larger resonators reduce the value of $C_p$. Therefore, the selected resonator size must account for the available pulse power and the feasibility of using the cryogenic LNA (see quantitative discussion in the Appendix). In our current design, we achieved a concentration sensitivity of ~2.6 μM/√Hz at 200 K for the bi-radical sample using the ParPar100_W resonator, which has a sample mode volume of ~0.3 nl. However, due to our limited bridge power, the need for a directional coupler to minimize external thermal noise, and the precautions taken to prevent potential LNA damage, we were limited to a $\pi$ pulse length of ~100–150 ns. This pulse length is too long for most DEER applications. Given these constraints, a more suitable resonator for such experiments would likely be the ParPar10_W, which offers a better balance between spin concentration sensitivity and $C_p$ values. In future work, this resonator could also be integrated with microfluidic capabilities to accommodate liquid samples [35].

In conclusion, cryogenic W-band ESR probeheads that combine cryogenic LNAs and surface micro-resonators, as described here, have the potential to become a valuable tool in ESR spectroscopy. They should enable the use of extremely small samples (sub-nanoliter) while benefiting from the advantages of high-field ESR, including enhanced spectral resolution and orientation selectivity, without compromising either absolute or concentration spin sensitivities. In fact, these designs may even improve upon these sensitivities compared to those at lower frequencies. However, additional advancements in sample preparation and placement on the resonators, as well as further improvements to the probehead (such as incorporating a low-loss limiter to protect the LNA) and spectrometer design, are necessary to maximize the method's utility. Based on our measured sensitivity values and the calculated data in Table 2, we anticipate a potential improvement of 2–3 orders of magnitude in absolute spin sensitivity if geometry-optimized samples are used, particularly with the smallest resonator (ParPar2_W). For instance, we could achieve a sensitivity of approximately 100 spins/√Hz for optimized samples (with $T_1 \approx T_2$) at ~10 K. These advancements could ultimately bring general-purpose induction-detection ESR techniques closer to achieving single-electron spin sensitivity in some samples, while maintaining reasonable averaging times of 1–10 hours.

## VI   Appendix: mm-wave excitation power and thermal noise considerations

As noted in the text, to fully leverage the benefits of the cryogenic LNA at low temperatures, the incoming mm-wave excitation signal must be attenuated to reduce the thermal noise originating from the room temperature source. For experiments that are not sensitive to excitation power, this does not pose a significant issue. However, for experiments such as DEER, which are often constrained by the available excitation power, this can introduce some challenges. In this section, we quantitatively evaluate the net gain in concentration sensitivity for a given experimental setup as a function of the applied attenuation.

Let us assume that we have a directional coupler or another type of cold attenuator that reduces the incoming noise power (and consequently, the pulse excitation power) by a factor of $dB$ dB. We also assume that the sample, the attenuator, and the LNA are all at temperature $T$, with the LNA contributing negligible noise. In this case, the noise temperature after the attenuator is:

$$T_{noise} = 10^{-dB/10} \times 300 + (1 - 10^{-dB/10}) \times T , \qquad (A.1)$$

This reduction in noise will improve the SNR for the ESR experiment. However, for experiments like DEER, the attenuation will also reduce the bandwidth of excitation (proportional to the mm-wave magnetic field, $B_1$), by a factor of $\sqrt{10^{-dB/10}}$, which would reduce the modulation depth, $\lambda$, of the DEER curve in a similar manner. This would dimmish some of the advantages of using the LNA. Overall, we can therefore write the improvement in the "usable" signal per noise ratio in his case as:

$$SNR_{improve}(dB, T) = \left( \frac{600}{T + (10^{-dB/10} \times 300 + (1 - 10^{-dB/10}) \times T)} \times \sqrt{10^{-dB/10}} \right)^{\frac{1}{2}}, \quad (A.2)$$

where we look at the ratio of the noise temperature of the LNA and the noise coming from the mm-wave excitation line, divided by the modulation depth without the attenuator (the latter is assumed to be 1 for convenience), to the same quantity with an attenuator. The assumption here is that the loss of signal reaching an LNA placed outside the cryostat is similar to the loss of signal to the LNA in the cryostat. Figure A1 shows the expected available SNR improvement as a function of the excitation power attenuation in dB, for several temperatures, based on eq (A.2). It is clear that attenuation of excitation power, which negatively affects the available contrast of the DEER experiment severely limits the usable SNR improvement by the used of the LNA. For example, while in ideal conditions one can expect that the noise at 4 K would be $\sim\sqrt{(300/4)}\sim8.7$ times better than at 30 K, our calculation shows that the possible effective improvement is $\sim2.5$; and at 50 K the expected sensitivity gain would drop from $\sim3.5$ to $\sim1.3$. Nevertheless, if the loss of signal leading to an external LNA is large, as noted in the Introduction, there is additional incentive for using the cryogenic LNA that adds to the net gain of the usable SNR.

One possible way to mitigate the constraint imposed by the cold attenuator is by reducing the size of the resonator, as the mm-wave magnetic field component for a given mm-wave power scales inversely with the square of the resonator volume. For instance, a 6 dB power reduction, which decreases $B_1$ by a factor of 2, could be compensated for by reducing the resonator volume by a factor of 4. However, according to Eq. (1), this would result in a reduction in concentration sensitivity by a factor of 2, effectively canceling out any potential gain in such sensitivity values. Additionally, smaller resonators generally have a lower internal $Q$ compared to larger resonators, which would further reduce sensitivity.

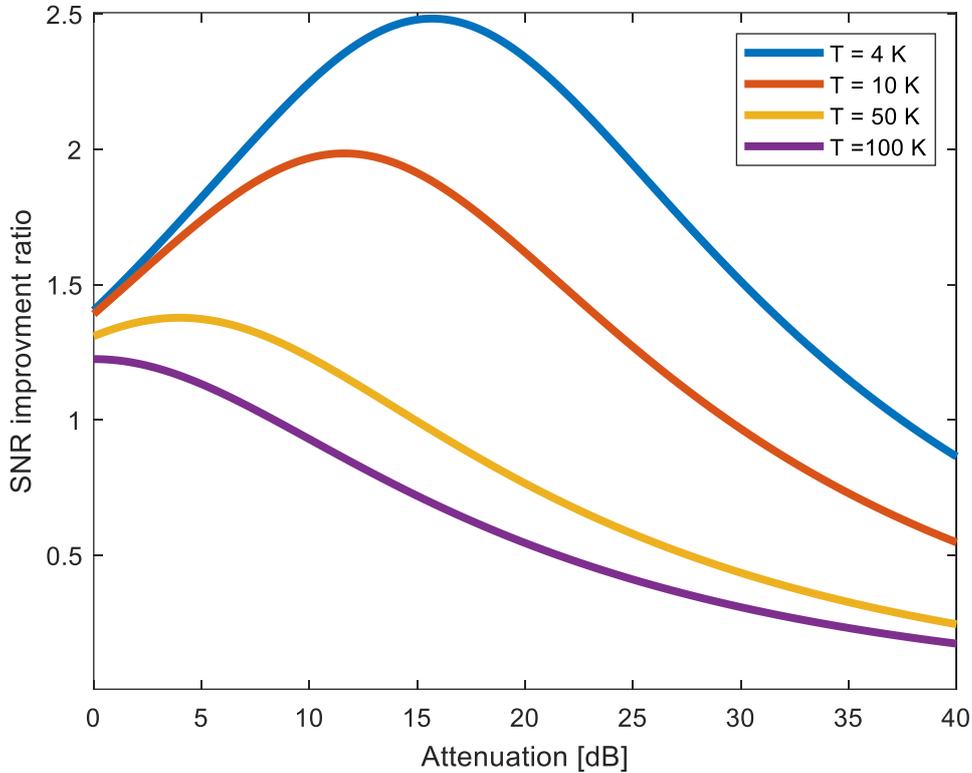

**Figure A1:** **Theoretical "usable" SNR improvement when using LNA and cold attenuator or directional coupler for the case of power limited ESR signal.**

## VII  Acknowledgements

This work was partially supported by grant #1357/21 from the Israel Science Foundation (ISF) and grant #FA9550-13-1-0207 from the Air Force Office of Scientific Research (AFOSR). We gratefully acknowledge the contributions of Prof. Jack Freed and Mr. Curt Dunnam from Cornell University during the initial development phase under the AFOSR award. The research was also supported by grant #3-12372 from the Israeli Ministry of Science. Additionally, this joint research project received financial support from the state of Lower Saxony and the Volkswagen Foundation, Hannover,

# VIII  References

[1] A. Blank, Y. Twig, Y. Ishay, Recent trends in high spin sensitivity magnetic resonance, J. Magn. Reson., 280 (2017) 20-29.

[2] D. Rugar, R. Budakian, H.J. Mamin, B.W. Chui, Single spin detection by magnetic resonance force microscopy, Nature, 430 (2004) 329-332.

[3] Y. Manassen, R.J. Hamers, J.E. Demuth, A.J. Castellano, Direct Observation of the Precession of Individual Paramagnetic Spins on Oxidized Silicon Surfaces, Phys. Rev. Lett., 62 (1989) 2531-2534.

[4] C. Durkan, M.E. Welland, Electronic spin detection in molecules using scanning-tunneling-microscopy-assisted electron-spin resonance, Appl. Phys. Lett., 80 (2002) 458-460.

[5] A. Morello, J.J. Pla, F.A. Zwanenburg, K.W. Chan, K.Y. Tan, H. Huebl, M. Mottonen, C.D. Nugroho, C.Y. Yang, J.A. van Donkelaar, A.D.C. Alves, D.N. Jamieson, C.C. Escott, L.C.L. Hollenberg, R.G. Clark, A.S. Dzurak, Single-shot readout of an electron spin in silicon, Nature, 467 (2010) 687-691.

[6] D. Vasyukov, Y. Anahory, L. Embon, D. Halbertal, J. Cuppens, L. Neeman, A. Finkler, Y. Segev, Y. Myasoedov, M.L. Rappaport, M.E. Huber, E. Zeldov, A scanning superconducting quantum interference device with single electron spin sensitivity, Nat Nanotechnol, 8 (2013) 639-644.

[7] J. Wrachtrup, C. Vonborczyskowski, J. Bernard, M. Orrit, R. Brown, Optical-Detection of Magnetic-Resonance in a Single Molecule, Nature, 363 (1993) 244-245.

[8] F. Jelezko, J. Wrachtrup, Single defect centres in diamond: A review, Physica Status Solidi a-Applications and Materials Science, 203 (2006) 3207-3225.

[9] B. Grotz, J. Beck, P. Neumann, B. Naydenov, R. Reuter, F. Reinhard, F. Jelezko, J. Wrachtrup, D. Schweinfurth, B. Sarkar, P. Hemmer, Sensing external spins with nitrogen-vacancy diamond, New J Phys, 13 (2011) 055004.

[10] J. Wrachtrup, A. Finkler, Single spin magnetic resonance, J. Magn. Reson., 269 (2016) 225-236.

[11] A. Blank, Spectroscopy, imaging, and selective addressing of dark spins at the nanoscale with optically detected magnetic resonance, Physica Status Solidi B-Basic Solid State Physics, 253 (2016) 1167-1176.

[12] M. Gulka, D. Wirtitsch, V. Ivády, J. Vodnik, J. Hruby, G. Magchiels, E. Bourgeois, A. Gali, M. Trupke, M. Nesladek, Room-temperature control and electrical readout of individual nitrogen-vacancy nuclear spins, Nat Commun, 12 (2021) 4421.

[13] Z. Wang, L. Balembois, M. Rančić, E. Billaud, M. Le Dantec, A. Ferrier, P. Goldner, S. Bertaina, T. Chanelière, D. Esteve, D. Vion, P. Bertet, E. Flurin, Single-electron spin resonance detection by microwave photon counting, Nature, 619 (2023) 276-281.

[14] P.P. Borbat, R.H. Crepeau, J.H. Freed, Multifrequency two-dimensional Fourier transform ESR: An X/Ku-band spectrometer, J. Magn. Reson., 127 (1997) 155-167.

[15] W.B. Mims, Electron Spin Echoes, in: G. S. (Ed.) Electron Paramagnetic Resonance, Plenum Press, New York,, 1972, pp. 263-351.

[16] G.A. Rinard, R.W. Quine, J.R. Harbridge, R.T. Song, G.R. Eaton, S.S. Eaton, Frequency dependence of EPR signal-to-noise, J. Magn. Reson., 140 (1999) 218-227.

[17] A. Blank, J.H. Freed, ESR microscopy and nanoscopy with "Induction" detection, Isr. J. Chem., 46 (2006) 423-438.

[18] Y. Twig, A. Sorkin, D. Cristea, A. Feintuch, A. Blank, Surface loop-gap resonators for electron spin resonance at W-band, Rev. Sci. Instrum., 88 (2017).

[19] R. Narkowicz, D. Suter, I. Niemeyer, Scaling of sensitivity and efficiency in planar microresonators for electron spin resonance, Rev. Sci. Instrum., 79 (2008) 084702.

[20] Y. Twig, E. Dikarov, A. Blank, Ultra miniature resonators for electron spin resonance: Sensitivity analysis, design and construction methods, and potential applications, Mol. Phys., 111 (2013) 2674-2682.

[21] Y. Artzi, Y. Twig, A. Blank, Induction-detection electron spin resonance with spin sensitivity of a few tens of spins, Appl. Phys. Lett., 106 (2015) 084104.

[22] C. Eichler, A.J. Sigillito, S.A. Lyon, J.R. Petta, Electron Spin Resonance at the Level of 10(4) Spins Using Low Impedance Superconducting Resonators, Phys. Rev. Lett., 118 (2017) 037701.

[23] A. Bienfait, J.J. Pla, Y. Kubo, M. Stern, X. Zhou, C.C. Lo, C.D. Weis, T. Schenkel, M.L.W. Thewalt, D. Vion, D. Esteve, B. Julsgaard, K. Molmer, J.J.L. Morton, P. Bertet, Reaching the quantum limit of sensitivity in electron spin resonance, Nat Nanotechnol, 11 (2016) 253-257.

[24] O.W.B. Benningshof, H.R. Mohebbi, I.A.J. Taminiau, G.X. Miao, D.G. Cory, Superconducting microstrip resonator for pulsed ESR of thin films, J. Magn. Reson., 230 (2013) 84-87.

[25] S. Vasilyev, J. Järvinen, E. Tjukanoff, A. Kharitonov, S. Jaakkola, Cryogenic 2 mm wave electron spin resonance spectrometer with application to atomic hydrogen gas below 100 mK, Rev. Sci. Instrum., 75 (2004) 94-98.

[26] K. Möbius, A. Savitsky, High-field/High-frequency EPR Spectroscopy in Protein Research: Principles and Examples, Appl Magn Reson, 54 (2023) 207-287.

[27] V. Kalendra, J. Turčak, J. Banys, J.J.L. Morton, M. Šimėnas, X- and Q-band EPR with cryogenic amplifiers independent of sample temperature, J. Magn. Reson., 346 (2023) 107356.

[28] V. Ranjan, S. Probst, B. Albanese, T. Schenkel, D. Vion, D. Esteve, J.J.L. Morton, P. Bertet, Electron spin resonance spectroscopy with femtoliter detection volume, Appl. Phys. Lett., 116 (2020) 184002.

[29] W.J. Wallace, R.H. Silsbee, Microstrip resonators for electron-spin resonance, Rev. Sci. Instrum., 62 (1991) 1754-1766.

[30] H.E. Altink, T. Gregorkiewicz, C.A.J. Ammerlaan, Sensitive electron paramagnetic resonance spectrometer for studying defects in semiconductors, Rev. Sci. Instrum., 63 (1992) 5742-5749.

[31] S. Pfenninger, W. Froncisz, J.S. Hyde, Noise Analysis of EPR Spectrometers with Cryogenic Microwave Preamplifiers, Journal of Magnetic Resonance, Series A, 113 (1995) 32-39.

[32] G.A. Rinard, R.W. Quine, R.T. Song, G.R. Eaton, S.S. Eaton, Absolute EPR spin echo and noise intensities, J. Magn. Reson., 140 (1999) 69-83.

[33] A. Blank, E. Dikarov, R. Shklyar, Y. Twig, Induction-detection electron spin resonance with sensitivity of 1000 spins: En route to scalable quantum computations, Phys. Lett. A, 377 (2013) 1937-1942.

[34] R. Narkowicz, H. Ogata, E. Reijerse, D. Suter, A cryogenic receiver for EPR, J. Magn. Reson., 237 (2013) 79-84.

[35] N. Dayan, Y. Ishay, Y. Artzi, D. Cristea, B. Driesschaert, A. Blank, Electron spin resonance microfluidics with subnanoliter liquid samples, Journal of Magnetic Resonance Open, 2-3 (2020) 100005.

[36] N. Dayan, Y. Ishay, Y. Artzi, D. Cristea, E. Reijerse, P. Kuppusamy, A. Blank, Advanced surface resonators for electron spin resonance of single microcrystals, Rev. Sci. Instrum., 89 (2018) 124707.

[37] J.S. Hyde, W. Froncisz, J.W. Sidabras, T.G. Camenisch, J.R. Anderson, R.A. Strangeway, Microwave frequency modulation in CW EPR at W-band using a loop-gap resonator, J. Magn. Reson., 185 (2007) 259-263.

[38] D.E. Budil, K.A. Earle, Sample Resonators for Quasioptical EPR, in: O.Y. Grinberg, L.J. Berliner (Eds.) Very High Frequency (VHF) ESR/EPR, Springer US, Boston, MA, 2004, pp. 353-399.

[39] S. Milikisiyants, A.A. Nevzorov, A.I. Smirnov, Photonic band-gap resonators for high-field/high-frequency EPR of microliter-volume liquid aqueous samples, J. Magn. Reson., 296 (2018) 152-164.

[40] I. Stil, A.L. Fontana, B. Lefranc, A. Navarrini, P. Serres, K.F. Schuster, Loss of WR10 Waveguide across 70-116 GHz, in: 22nd International Symposium on Space Terahertz Technology, Tokyo, 2012.

[41] G.A. Ediss, Measurements and Simulations of Overmoded Waveguide Components at 70-118 GHz, 220-330 GHz and 610-720 GHz, in: 14th International Symposium on Space Ter-ahem: Technology, 2004.

[42] Y. Twig, E. Suhovoy, A. Blank, Sensitive surface loop-gap microresonators for electron spin resonance, Rev. Sci. Instrum., 81 (2010) 104703.

[43] H. Malissa, D.I. Schuster, A.M. Tyryshkin, A.A. Houck, S.A. Lyon, Superconducting coplanar waveguide resonators for low temperature pulsed electron spin resonance spectroscopy, Rev. Sci. Instrum., 84 (2013).

[44] A.V. Matheoud, G. Gualco, M. Jeong, I. Zivkovic, J. Brugger, H.M. Ronnow, J. Anders, G. Boero, Single-chip electron spin resonance detectors operating at 50 GHz, 92 GHz, and 146 GHz, J. Magn. Reson., 278 (2017) 113-121.

[45] S. Probst, A. Bienfait, P. Campagne-Ibarcq, J.J. Pla, B. Albanese, J.F.D. Barbosa, T. Schenkel, D. Vion, D. Esteve, K. Molmer, J.J.L. Morton, R. Heeres, P. Bertet, Inductive-detection electron-spin resonance spectroscopy with 65 spins/root Hz sensitivity, Appl. Phys. Lett., 111 (2017).

[46] N. Abhyankar, A. Agrawal, P. Shrestha, R. Maier, R.D. McMichael, J. Campbell, V. Szalai, Scalable microresonators for room-temperature detection of electron spin resonance from dilute, sub-nanoliter volume solids, Science Advances, 6 (2020) eabb0620.

[47] T. Yalcin, G. Boero, Single-chip detector for electron spin resonance spectroscopy, Rev. Sci. Instrum., 79 (2008).

[48] J. Anders, A. Angerhofer, G. Boer, K-band single-chip electron spin resonance detector, J. Magn. Reson., 217 (2012) 19-26.

[49] J.P. Campbell, J.T. Ryan, P.R. Shrestha, Z.L. Liu, C. Vaz, J.H. Kim, V. Georgiou, K.P. Cheung, Electron Spin Resonance Scanning Probe Spectroscopy for Ultrasensitive Biochemical Studies, Anal. Chem., 87 (2015) 4910-4916.

[50] J.W. Sidabras, J. Duan, M. Winkler, T. Happe, R. Hussein, A. Zouni, D. Suter, A. Schnegg, W. Lubitz, E.J. Reijerse, Extending electron paramagnetic resonance to nanoliter volume protein single crystals using a self-resonant microhelix, Science Advances, 5 (2019) eaay1394.

[51] C. Bonizzoni, M. Maksutoglu, A. Ghirri, J. van Tol, B. Rameev, M. Affronte, Coupling Sub-nanoliter BDPA Organic Radical Spin Ensembles with YBCO Inverse Anapole Resonators, Appl Magn Reson, 54 (2023) 143-164.

[52] B. Pietzak, M. Waiblinger, T.A. Murphy, A. Weidinger, M. Höhne, E. Dietel, A. Hirsch, Buckminsterfullerene C60: a chemical Faraday cage for atomic nitrogen, Chem. Phys. Lett., 279 (1997) 259-263.

[53] W.V. Smith, P.P. Sorokin, I.L. Gelles, G.J. Lasher, Electron-Spin Resonance of Nitrogen Donors in Diamond, Physical Review, 115 (1959) 1546-1552.

[54] S. Knorr, A. Grupp, M. Mehring, M. Waiblinger, A. Weidinger, Electron spin relaxation rates T1−1 and T2−1 in diluted solid N@C60, AIP Conf. Proc., 544 (2000) 191-194.

[55] R.M. Brown, A.M. Tyryshkin, K. Porfyrakis, E.M. Gauger, B.W. Lovett, A. Ardavan, S.A. Lyon, G.A.D. Briggs, J.J.L. Morton, Coherent State Transfer between an Electron and Nuclear Spin in $^{\mathbf{15}}\mathbf{N}\mathbf{@}{\mathbf{C}}_{\mathbf{60}}$, Phys. Rev. Lett., 106 (2011) 110504.

[56] G. Mitrikas, Encapsulated Atomic Hydrogen in Octamethyl-POSS Cages: A Pulsed EPR Study, ChemPlusChem, n/a  e202400146.

[57] D. Goldfarb, Y. Lipkin, A. Potapov, Y. Gorodetsky, B. Epel, A.M. Raitsimring, M. Radoul, I. Kaminker, HYSCORE and DEER with an upgraded 95GHz pulse EPR spectrometer, J. Magn. Reson., 194 (2008) 8-15.

[58] W. Hofbauer, K.A. Earle, C.R. Dunnam, J.K. Moscicki, J.H. Freed, High-power 95 GHz pulsed electron spin resonance spectrometer, Rev. Sci. Instrum., 75 (2004) 1194-1208.

[59] F.H. Cho, V. Stepanov, S. Takahashi, A high-frequency electron paramagnetic resonance spectrometer for multi-dimensional, multi-frequency, and multi-phase pulsed measurements, Rev. Sci. Instrum., 85 (2014) 075110.

[60] E. Dikarov, M. Fehr, A. Schnegg, K. Lips, A. Blank, Selective electron spin resonance measurements of micrometer-scale thin samples on a substrate, Meas. Sci. Technol., 24 (2013) 115009.